\documentclass[10pt,journal,compsoc]{IEEEtran}

\ifCLASSOPTIONcompsoc
  \usepackage[nocompress]{cite}
\else
  \usepackage{cite}
\fi
\usepackage{caption}
\usepackage{xcolor}
\usepackage{multirow}
\ifCLASSINFOpdf
   \usepackage[pdftex]{graphicx}
\else
\fi
\begin{document}
%
\title{AffectiveNet: Affective-Motion Feature Learning for Micro Expression Recognition}
%
%
%
%

\author{Monu~Verma,~\IEEEmembership{Student,~Member,~IEEE,}
        Santosh~Kumar~Vipparthi,~\IEEEmembership{Member,~IEEE,}
        and~Girdhari~Singh
\IEEEcompsocitemizethanks{\IEEEcompsocthanksitem Monu Verma, Santosh Kumar Vipparthi and Girdhari Singh are with Vision Intelligence Lab at Department of Computer Science and Engineering, Malaviya National Institute of Technology, Jaipur, India (Email: monuverma.cv@gmail.com;  skvipparthi@mnit.ac.in; gsingh.cse@mnit.ac.in)\protect\\
}
}

%
%

\markboth{IEEE MultiMedia,~Vol.~xx, No.~xx, August~2020}%
{Shell \MakeLowercase{\textit{et al.}}: Bare Advanced Demo of IEEEtran.cls for IEEE Computer Society Journals}
%



\IEEEtitleabstractindextext{%
\begin{abstract}
Micro-expressions are hard to spot due to fleeting and involuntary moments of facial muscles. Interpretation of micro emotions from video clips is a challenging task. In this paper we propose an affective-motion imaging that cumulates rapid and short-lived variational information of micro expressions into a single response. Moreover, we have proposed an AffectiveNet: affective-motion feature learning network that can perceive subtle changes and learns the most discriminative dynamic features to describe the emotion classes. The AffectiveNet holds two blocks: MICRoFeat and MFL block. MICRoFeat block conserves the scale-invariant features, which allows network to capture both coarse and tiny edge variations. While MFL block learns micro-level dynamic variations from two different intermediate convolutional layers. Effectiveness of the proposed network is tested over four datasets by using two experimental setups: person independent (PI) and cross dataset (CD) validation. The experimental results of the proposed network outperforms the state-of-the-art approaches with significant margin for MER approaches.
\end{abstract}

}

\maketitle

\IEEEdisplaynontitleabstractindextext

%
\IEEEpeerreviewmaketitle

\ifCLASSOPTIONcompsoc
\IEEEraisesectionheading{\section{Introduction}\label{sec:introduction}}
\else
\section{Introduction}
\label{sec:introduction}
\fi

%
%
%
%
\IEEEPARstart{M}{icro}  expressions (ME) have rich source of information to reveal the true emotions of a person. MEs  originate in high stake situations, when a person is trying to repress his/her genuine feelings within manifested expressions (macro expressions). Usually, macro expression active on a face for 4 to 5 seconds that can be perceived easily. whereas, MEs are active for 1/30 to 1/25 seconds. Since MEs are short-lived and fleeting in nature, it is hard to differentiate them through naked eyes. Earlier, trained persons were able to spot micro-expressions but achieve less than 50$\%$ accuracy. Analysis of real emotions from video clips has a wide range of applications such as: depression analysis, police interrogation, law enforcement, multi-media entertainment, clinical diagnosis etc.

In literature many feature extraction methods (handcrafted / traditional feature descriptors) were proposed as spatiotemporal LBP with integral projection (STLBP-IP) \cite{stLbp} and FDM \cite{fdm} to encode spatial and temporal changes from the ME video sequences. However, handcrafted feature descriptors were focused only on superficial features and failed to capture sufficient features of MEs.

Nowadays, with the advent of technology, deep learning models \cite{vgg,resnet,mobilenet} have gained popularity in solving various computer vision tasks like image classification, semantic segmentation, face authorization, biometric, and many more.  Recently, some deep models \cite{microatt,microlstm,spontenous,spotrcnn} are proposed to deal with micro-expression recognition. However, most of the existing methods use combination of CNN and RNN or CNN and LSTM to extract the spatial and temporal features, respectively. These methods first capture the spatial features from the frames and then these are fed to RNN or LSTM to fetch the temporal features. Thereby these approaches failed to establish relationship between spatial and temporal features occurring simultaneously in frames and lead to degrading the performance.

Inspired from the literature \cite{resnet,mobilenet}, a novel affective-motion feature learning method is proposed to learn and classify the features of micro expressions. The AffectiveNet has ability to capture spatial and temporal features simultaneously from the affective-motion images. The contributions of the proposed approach are summarized as follows:
\begin{figure*}[!t]
\centering
	\includegraphics[width=\linewidth,height=3.0in]{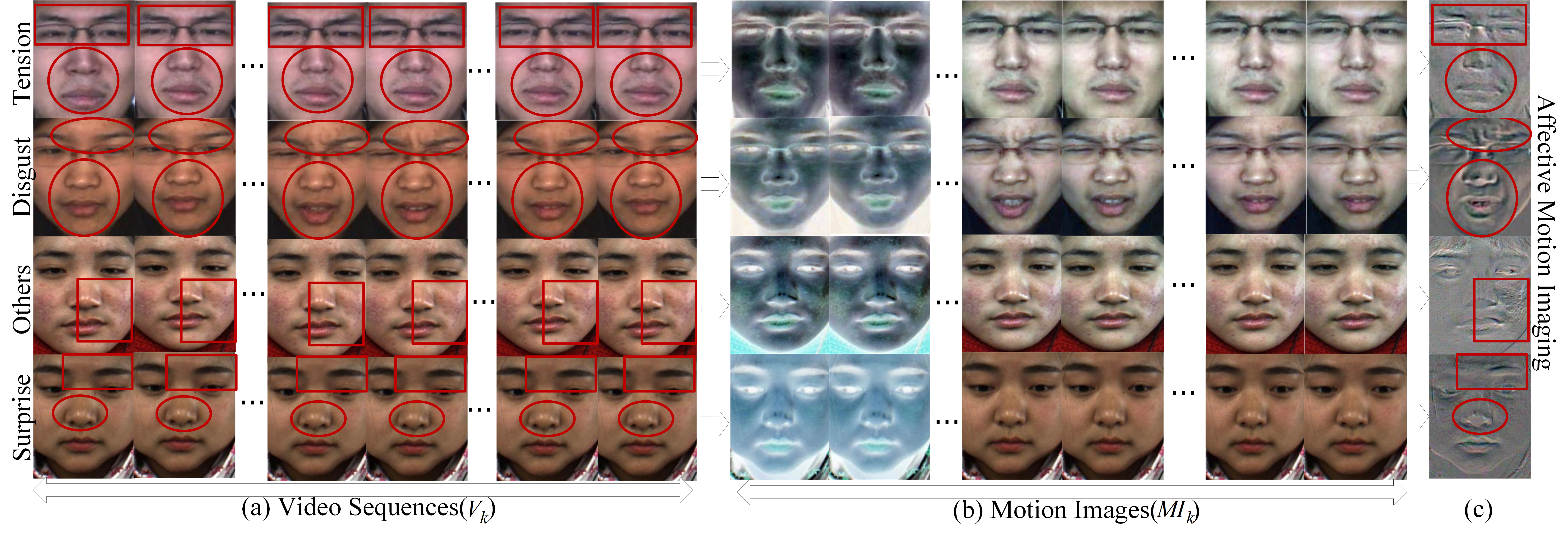}
	\caption{Visualization of (a) input video (V) with k frames, (b) Motion Images (MI) generated by multiplying input frames with coefficients Frame weights (Fw) and (c) Affective-Motion Images (AMI) representing both appearance and motion that occurred between the frames in a 2d image.}
	\label{fig:Figure1}
\end{figure*}
\begin{enumerate}
\item We propose an affective-motion imaging that summarizes the spatial structure features with temporal variations into one image instance.
\item We propose an AffectiveNet for micro-expression recognition by introducing two blocks: MICRoFeat and MFL blocks. MICRoFeat block has been proposed to enhance  learning capability of the network by  capturing multi-scale features.MFL block has been designed to increase the discriminability of the network as it is able to learn micro-level features. 
\item The effectiveness of the proposed AffcetiveNet is examined by adopting two validation schemes: person independent and cross dataset over for benchmark datasets and compared with state-of-the-art MER approaches. 

\end{enumerate}

\section{Literature Review}
Feature extraction is an essential part of MER task. Wang et al. \cite{tics} introduced a tensor independent color space model (TICS) by representing the image sequences in 4D structure such as: 2D structure represents the spatial texture patterns, 3rd dimension for momentary variation features and 4th dimension describes RGB color components to spot the micro-expressions.  Furthermore, they extended thier work and proposed sparse tensor canonical correlation method \cite{sparse} to analyse the micro expressions movements. Happy et al. \cite{fhofo} proposed a fuzzy histogram-based optical flow orientation technique (FHOFO) to capture temporal features of the micro-expressions. Wang et al. \cite{mdmd} introduced the main directional maximal difference (MDMD) to capture the facial expressive movements by extracting the maximal magnitude difference in between the optical flow directions.

Recently, the adoption of deep learning networks of VGG Net \cite{vgg}, ResNet \cite{resnet} and MobileNet \cite{mobilenet} have created a tremendous take-off in the field of computer vision. The literature on MER shows that, convolutional neural networks (CNN) based models also achieve impressive results up to some extent. 

Furthermore, an evolutionary search is being applied to detect the disparities between the frames of micro expressions. Wang et al. \cite{microatt} introduced a micro attention module in resnet \cite{resnet} that mainly focused on expressive regions which included most of the action units. To capture the action units they utilized the transfer learning from macro to micro expressions.khor et al. \cite{cnnlstm} adopted CNN network and long short-term memory (LSTM) to learn the Spatio-temporal information for each image frame. Wang et al. \cite{microlstm} utlized the CNN model for visual feature extraction and LSTM for sequence learning between the frames to spot the micro expressions. Moreover, Li et al. \cite{3dflow} introduced a 3d flow CNN network, which incorporated optical flow information with CNN network to learned deep features of minute variation responsible to spot micro expression class. xia et al.   \cite{spontenous} proposed a recurrent convoloution neural network to capture the features of subtle changes occurred between image sequences.Liong et al. \cite{ststsnet} also utilized optical flow to represents flow variations between frames and feed them to three parallelly connected CNN layer streams that learn the salient features of micro-expressions and classify them accordingly. Xia et al. \cite{spotrcnn}proposed a extended recurrent convolution network to extract the spatial-temporal deformations of
micro-expression sequence by considering appearance and geometrical information, respectively .
\begin{figure*}[!t]
\centering
\captionsetup{justification=centering}
	\includegraphics[width=\linewidth, height=2.8in]{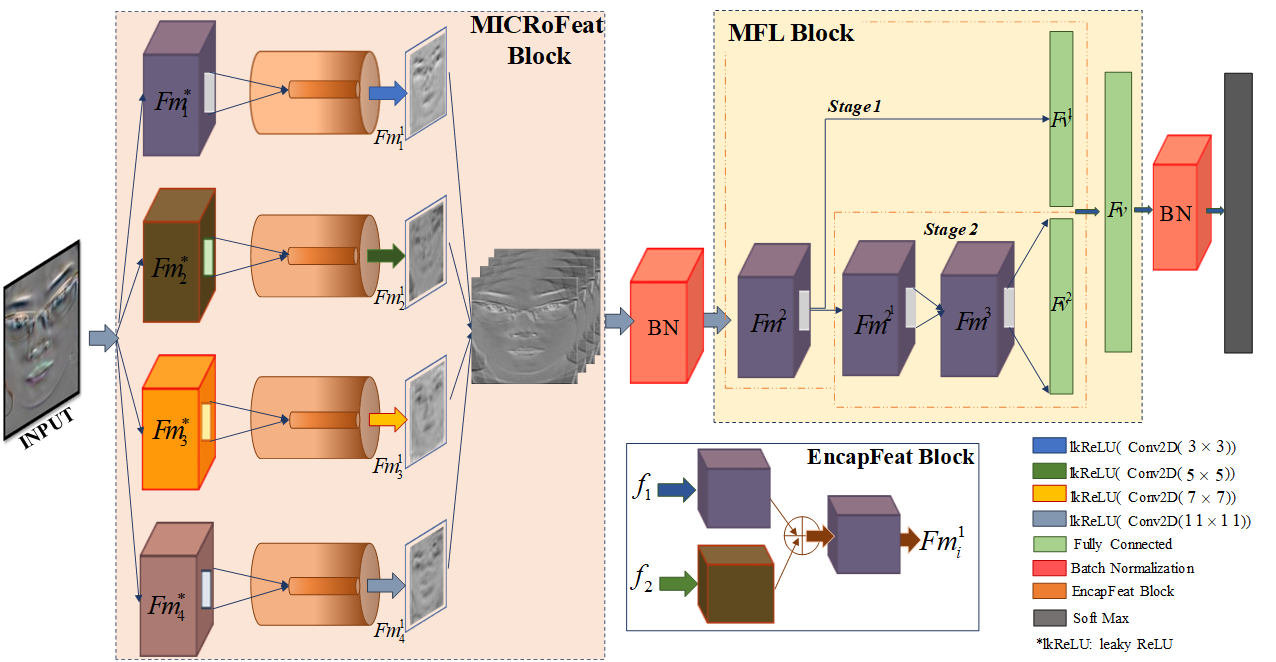}
	
	\vspace{-3mm}
	    \caption{The detailed architecture of the proposed Affective Network for micro expression recognition.}

	\label{fig:Figure2}
\end{figure*}

\begin{figure}[t]
\begin{center}
   \includegraphics[width=1\linewidth,height=4.0in]{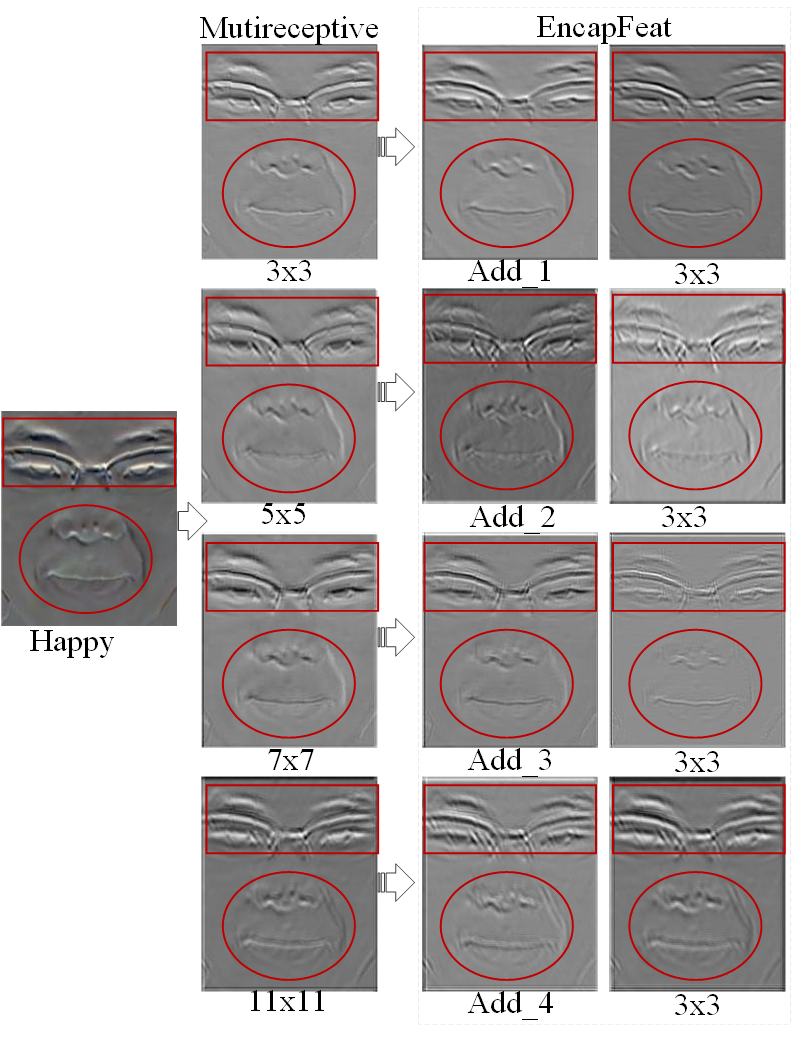}
\end{center}
\vspace{-3mm}
   \caption{The feature maps of happy image produced at each convolutional layer of MICRoFeat block.}
\label{fig:Figure3}
\end{figure}

\section{Proposed Method}
Micro-expressions appear only in the few frames of a video due to fleeting and short-lived nature. Therefore, interpretation of the content in a video and spotting micro-expressions between the frames is a challenging task. In literature the state-of-art MER systems apply complex algorithms to represent the adequate video content . Moreover, all benchmark datasets hold variant size video sequences, thus most of the state-of-the-art approaches utilized the time interpolation to normalize the dataset. It may lose or alter the domain knowledge of micro-expressions by shearing or filling holes in between the frames. To address these issues, in this paper we have proposed affective-motion imaging (AMI). Affective-motion image represents video content into a single instance by preserving high stake active dynamics of micro expressions. Hence, we have used an AffectiveNet learn the dynamics of micro-expressions and interprets the relevant emotion class.
\subsection{Affective-Motion Imaging}
Inspired from the literature \cite{dynamic} in this paper we  introduced affective-motion imaging  (AMI). AMI interprets the content of the video by focusing on the facial moving regions and compress that into a single instance. Therefore, affective-motion image implies movements in a still image by summarizing spatial and temporal dynamics of the whole video frames. To construct a single image instance from video sequences, we  estimate the motion between the frames and allocated ranks to video frames by using a ranking function. Let LR is a ranking function, which updates the Rank of frames by using Eq. \ref{eq:1}-\ref{eq:2}. 
\begin{equation}\label{eq:1}
    LR[1,i]=\frac{(2\times I[1,i])-k}{I[1,i]}
\end{equation}
\begin{equation}\label{eq:2}
    I[1,i]=[i,i+1,i+2,...k]
\end{equation}

where $I$ represents as index matrix with $i \in {1,2,..k}$ and $k$ implies the total number of frames extracted from the video . Furthermore, frame weight $Fw(i)$ is assigned to each frame by using Eq. \ref{eq:3}.
\begin{equation}\label{eq:3}
  Fw(i)=\sum_{j-1}^{k-i}{LR[1,j]}
\end{equation}

Moreover, motion images  are computed by utilizing Eq. \ref{eq:4}.
\begin{equation}\label{eq:4}
   MI_{i}=\nu_{i}\times Fw(i)
\end{equation}

where $\nu_{i} \in \nu$  represents the $i^{th}$ frame of the video $\nu$. Specifically, frame weights analyze the motions between the frames in a video and quantify it with the help of ranking function. Further, frames are amplified by multiplying frame weight coefficient named as motion images. Motion images magnify the temporal changes and abbreviated uniform information, as shown in Fig. 1. Finally, Affective-motion image is computed by merging all motion images such as. 
\begin{equation}\label{eq:5}
   AMI=\sum_{i}^{k}{MI_{i}}
\end{equation}

Samples of affective motion images are demonstrated in Fig.\ref{fig:Figure1}.  From Fig. \ref{fig:Figure1}, it is clearly visible that the affective motion images successfully preserve the influencing dynamics of micro-expressions within single frame. Moreover, affective motion images abbreviated uniform information and help to protrude nonuniform variations, highlighted in red blocks those play decision making role in MER. Further to learn effective features of micro-expressions affective motion images are forwarded to the AffectiveNet.
\begin{figure*}[!t]
	\includegraphics[width=1\linewidth, height=3.0in]{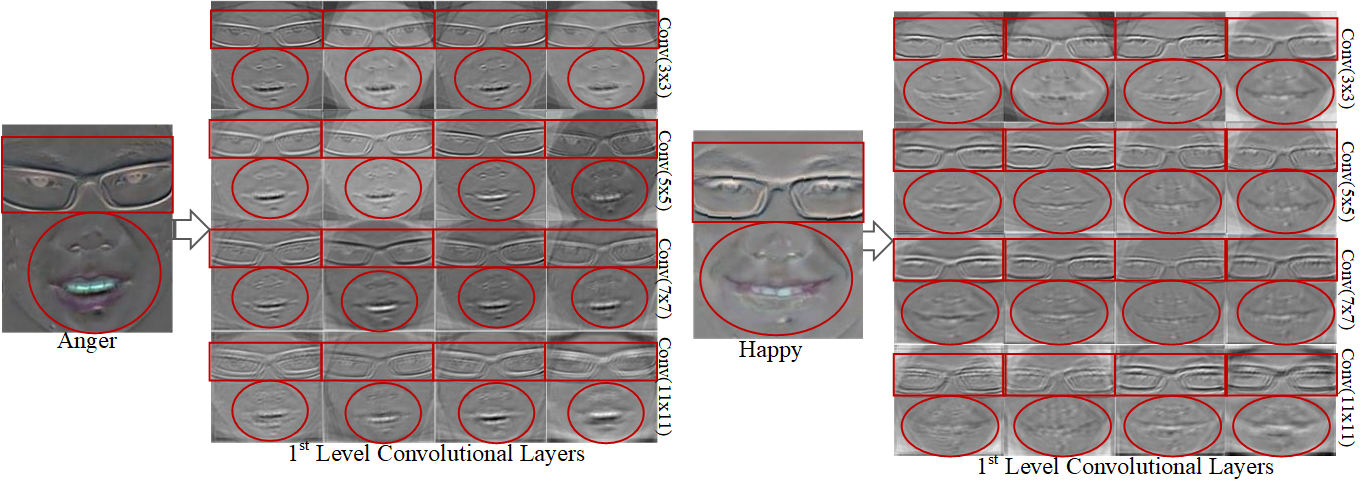}
	\vspace{-3mm}
	\caption{The feature maps generated from two emotion classes a) Anger and b) Happy, at 1st level convolutional layers of Affective Network. The region of interest (red block) shows that AffectiveNet is able to differentiate between two expression classes (inter-class).}
	\label{fig:Figure4}
\end{figure*}
\begin{table}
\centering
\caption{Recognition Accuracy Comparison on CASME-I and CSME-II Datasets. \textit{*This result is from the corresponding original paper and H, S, D, R, T, P, N, O, Sa, F stands for Happy, Surprise, Disgust, Repression, Tense, Positive, Negative, Others, Sad, Fear. AffectiveNet-2 represents results evaluated by following experimental setup used in STRCNN \cite{spotrcnn}}.}
\label{tab:Table1}
\vspace{-3mm}
\begin{tabular}{cccc}\hline
Method    & Task & CASME-I & CASME-II \\
\hline \hline
STLBP-IP*\cite{stLbp} &   $\left(H, S, D, R, O\right) $  &     N$/$A    & 59.91         \\
 FDM* \cite{fdm}         &    $\left(D, R, S, T\right)$   &   56.14      &    45.93      \\
  3D-Flow* \cite{3dflow}        &  $\left(H, S, D, R, T\right) $    & 55.44         &   59.11       \\
   TICS* \cite{tics}       &  $\left(P, N, S, O\right)$    &    61.86     & 61.11         \\
   FHOFO* \cite{fhofo}       & $\left(P, N, S, O\right)$     &    65.99     &    55.86      \\
  CNN-LSTM* \cite{cnnlstm}        &   $\left(H, S, D, R, O\right)$   & 60.98         &      N$/$A   
  \\
 MicroAtt* \cite{microatt}        &  $\left( A, D, F, H, Sa, S, O\right)$   & N/A         &     {65.90} 
  \\
  {Sp-RCNN* \cite{spontenous} }       &   {$\left(P, N, S, O\right)$}   & {63.20 }        &    {  65.80} 
  \\
 { STRCNN* \cite{spotrcnn}}        &   {$\left(P, N, S, O\right)$ }  & N$/$A        &      {56.00} 
  \\
    ResNet-50 \cite{resnet}      &    $\left(P, N, S, O\right)$  &  25.04       &     32.12     \\
   MobileNet \cite{mobilenet}        &   $\left(P, N, S, O\right)$   &      33.77   &       30.25   \\
  Af-Net-KS-1        &   \textbf{$\left(P, N, S, O\right)$}&     {\textbf{56.48}}    &  \textbf{45.64}          \\
    Af-Net-KS-2         &  \textbf{$\left(P, N, S, O\right)$}     &   {\textbf{60.26}}      &   \textbf{49.58}       \\
  Af-Net-LFC           &  \textbf{$\left(P, N, S, O\right)$}     &    { \textbf{56.51}}    &     \textbf{53.62}      \\
 {Af-Net-WoMFL }        &   {\textbf{$\left(P, N, S, O\right)$}}    & {   57.94 }   &    { \textbf{60.17}}      \\
 {Af-Net-$3\times3$ }         &   {\textbf{$\left(P, N, S, O\right)$}}    &   { \textbf{59.32} }    & { \textbf{54.12} }   \\
{Af-Net-$1\times1$  }        &   {\textbf{$\left(P, N, S, O\right)$}}    &   {\textbf{56.53}}     &  {\textbf{43.88}}        \\
AffectiveNet-1         &    \textbf{$\left(P, N, S, O\right)$}   &    \textbf{66.99}     &      \textbf{61.58}
\\
{AffectiveNet-2 }        &   {\textbf{$\left(P, N, S, O\right)$}}   &    {\textbf{72.64}}     &    {  \textbf{68.74}}
\\ \hline 
\vspace{-3mm}
\end{tabular}
\end{table}
\subsection{Affective Network}
In this paper we have proposed a portable CNN model affective motion feature learning (AffectiveNet) that learns the salient features of micro-expression by capturing momentary changes from the affective motion images. AffectiveNet mainly comprises of two blocks: multi-receptive feature preservative (MICRoFeat) block and microfeature learning (MFL) block as shown in Fig. \ref{fig:Figure2}.
\subsubsection{MICRoFeat Block}
Micro-level variations can be captured through affective-motion images, where the expressive regions may spread from small region to extensive regions. The micro-level expression variations are clearly depicted in Fig. \ref{fig:Figure1}. Although these changes are imperceptible but have a high impact in identifying the micro-expressions. Therefore, a robust CNN network that can elicit both coarse and detailed texture features are needed to acquire sufficient knowledge for adequate emotion classification. In literature, it has been confirmed that inferior variations like eyebrow lift, check crinkles, forehead wrinkles, glabella, chin, eyelid and lip lines can be captured through small sized convolutional filters, while the abstract changes like eyes, nose, mouth and lip shapes tend to respond with large sized filter. However, most of the CNN based models like VGG Net \cite{vgg}, ResNet \cite{resnet} and MobileNet \cite{mobilenet} hold uniform-sized filters. Thus, these networks degrade the performance of micro-expression recognition as they fail to acquire enough feature variations from affective-motion images. Therefore, in this paper, we have introduced MICRoFeat block to extract the detailed expressive features from the affective-motion images. The MICRoFeat block has ability to capture detailed features from small regions to extensive regions, by applying four convolutional (Conv) layers with multi-scale filters as $3\times3$, $5\times 5$, $7\times 7$ and $11\times11$. Let $I(u,v)$  be an input image and $\varepsilon_{S}^{x,N}\{\cdot \} $   represents conv function, where, $S$ implies for stride, $N$ is depth, $x$ stands for the size of filter. Then, output of Conv layers with multi-scale filters are computed by Eq. (\ref{eq:6}-\ref{eq:7}).
\begin{equation}\label{eq:6}
  Fm_{i}^{*}=\varepsilon_{1}^{p(i),16}\{I(u,v)\}
\end{equation}

where, $i={1,2,3,4}$, represents the each multi-scale conv layers and
\begin{equation}\label{eq:7}
  p(i)=[3,5,7,11]
   \end{equation}
   
Further, feature maps $Fm_{i}^{*}$ of each layer are forwarded to the next aligned encapsulated feature (EncapFeat) blocks. EncapFeat block imposes two Conv layers with different scales as $3\times3$ and $5\times 5$ to express the  edge variations of each muscle movement (those provokes facial expressions) by extracting coarse to fine edge variations. Furthermore, resultant feature maps are refined by employing  $3\times3$ Conv layer. Resultant feature maps $Fm_{i}^{1}$ of the EncapFeat are computed by using Eq.(\ref{eq:8}-\ref{eq:10}). 
\begin{equation}\label{eq:8}
  Fm_{i}^{1}=EncapFeat\{Fm_{i}^{*}\}
\end{equation}
\begin{equation}\label{eq:9}
  EncapFeat\{Fm_{i}^{*}\}=\varepsilon_{2}^{3,64}\{f_{1}\{Fm_{i}^{*}\}+f_{2}\{Fm_{i}^{*}\}\}
\end{equation}
\begin{equation}\label{eq:10}
  f_{k}\{Fm_{i}^{*}\}=\varepsilon_{2}^{2k+1,32}\{Fm_{i}^{*}\}
\end{equation}

Moreover, response of each EncapFeat block is coupled and forwarded to next down-sampled Conv layers such as.
\begin{equation}\label{eq:11}
  MICRoFeat=\{Fm_{1}^{1}\|Fm_{2}^{1}\|Fm_{3}^{1}\|Fm_{4}^{1}\}
\end{equation} 

where, $\|$, represents the concat operation.
The effectiveness of the MICRoFeat block is depicted in Fig. \ref{fig:Figure3}, where red highlighted boxes represent the expressive regions. From the Fig. \ref{fig:Figure3}, it is clear that small $\left(3\times3 , 5\times 5\right)$ and large $\left(7\times7 , 11\times 11\right)$ sized filters  are able to extract minute and high-level variations, respectively. 

\subsubsection{MFL block}
Inspired from the residual concept \cite{resnet}, we introduced a MFL module. The main aim of this module is to refine the micro-expression regions in two parallel stages as shown in the Fig.\ref{fig:Figure2}. In stage 1, low-level features of widespread expressive regions are learned through one FC layer. These low-level features increase the learning capability of the network. Similarly, in stage 2, micro variation in the high-level feature are forwarded parallelly to FC network. Further, resultant features of laterally connected FC layers are fused to capture micro-level variations in an expressive region. The lower layer features are effective for identifying variations in small regions. Thus, fusion of these features improve the discriminable capability between the inter and intra-class variations. Moreover, MFL block increases the learning capability with minimum number of parameters for AffectiveNet as compared to existing state-of-the-art approaches. Let FC represents the fully connected layer and concat implies for the depth concatenation function. Then, output feature vector $Fv$ is computed by Eq. (\ref{eq:12}-\ref{eq:16}).
\begin{equation}\label{eq:12}
  Fv=FC^{4}\left(\beta \{Fv^{1}\|Fv^{2}\} \right)
\end{equation}
\begin{equation}\label{eq:13}
   Fv^{1}=FC^{32}\left(Fm^{2}\right)
\end{equation}
\begin{equation}\label{eq:14}
   Fv^{2}=FC^{32}\left(Fm^{3}\right)
\end{equation}
\begin{equation}\label{eq:15}
   Fm^{2}=\varepsilon_{2}^{3,184}\{\beta\left(MICRoFeat\right)\}
\end{equation}
\begin{equation}\label{eq:16}
   Fm^{3}=\varepsilon_{2}^{3,196}\{\varepsilon_{2}^{3,128\{Fm^{2}\}}\}
\end{equation}

Where, $\beta$  represents the batch normalization (BN) function. BN is incorporated in proposed network to deal with the issue of divergence in feature distribution that occurs due to disproportion of image sets: training and testing data. Thereby, BN improves the strength of AffectiveNet by normalizing the feature responses of the preceding contact layer by subtracting batch mean and dividing by the standard deviation as follows: 
\begin{equation}\label{eq:17}
    p_{k}=\omega \bar {q_{k}}+\phi \beta\left(\bar{q_{k}}\right)
\end{equation}
\begin{equation}\label{eq:18}
    \bar{q}=\frac{q_{k} - n_{B}}{\sqrt{S_{B}^{2}+\epsilon}}
\end{equation}

where,$q_{k}$  is the mini-batch size  and $B=\{q_{1}, q_{2},..., q_{N}\}$  are the learnable parameters. $\omega$  and $S-{B}$ implies the mean and standard deviation of the batch as calculated using Eq. (\ref{eq:19}-\ref{eq:20}). 
\begin{equation}\label{eq:19}
    n_{B}=\frac{1}{N} \sum_{k=1}^{N}\left(q_{k}\right)
\end{equation}
\begin{equation}\label{eq:20}
     S_{B}=\frac{1}{N} \sum_{k=1}^{N}\left(q_{i}-n_{B}\right)^{2}
\end{equation}

The capability of the AffectiveNet to control intra-class (anger and happy) variations is depicted in Fig. \ref{fig:Figure4}. Thus, we can conclude that, proposed model is able to differentiate between two expression classes (inter-class variation) within an active patch. Active patches are highlighted by red color boxes.

\begin{table*}[]
\caption{Recognition Accuracy Comparision on CASME$^{2}$, SAMM AND CI2CII, CI2C$^{2}$, CI2S, CII2CI, CII2C$^{2}$, CII2S, S2CI, S2CII, S2C$^{2}$ for PIE and CDE Experiments, Respectively. \textit{Here, PIE, CDE and CSM are stands for Person Independent Experiment, Cross Data Experiments and CASME, respectively.}}
\label{tab:Table2}
\vspace{-3mm}
\begin{tabular}{|c|c|c|c||c|c|c||c|c|c||c|c|c||}
\hline
\multirow{3}{*}{\textbf{Method}}      & \textbf{Exp.}         & \multicolumn{2}{c||}{\textbf{PIE}}                                              & \multicolumn{9}{c||}{\textbf{CDE}}                                                                                                               \\ \cline{2-13} 
                             & \textbf{Training}     & \multirow{2}{*}{CASME$^{2}$} & \multirow{2}{*}{SAMM} & \multicolumn{3}{c||}{CASME-I}                & \multicolumn{3}{c||}{CASME-II}              & \multicolumn{3}{c||}{SAMM}                   \\ \cline{2-2} \cline{5-13} 
                             & \textbf{Testing}      &                                               &                       & CSM-II & CSM$^{2}$ & SAMM  & CSM-I & CSM$^{2}$ & SAMM  & CSM-I & CSM-II & CSM$^{2}$ 
                             \\ \hline
\multicolumn{2}{|l|}{{MicroAtt\cite{microatt}}}             & N/A                                         &{48.50}                 & N/A  & N/A                      & N/A & N/A & N/A                      & N/A  & N/A & N/A  & N/A                     

\\ \hline
\multicolumn{2}{|l|}{{STRCNN\cite{spotrcnn}}}             & N/A                                        & {54.45}                 & N/A  & N/A                      & N/A & N/A & N/A                      & N/A  & N/A & N/A  & N/A  
                             \\ \hline
\multicolumn{2}{|l|}{ResNet-50\cite{resnet}}             & 46.58                                         & 36.50                 & 10.67  & 12.50                      & 10.69 & 10.81 & 23.54                      & 8.81  & 20.15 & 16.60  & 29.06                     
              
\\ \hline
\multicolumn{2}{|l|}{MobileNet \cite{mobilenet}}             & 35.05                                         & 40.20                 & 11.86  & 4.00                       & 14.46 & 23.78 & 6.40                       & 18.24 & 14.05 & 12.64  & 8.00                       \\ \hline
\multicolumn{2}{|l|}{\textbf{AffectiveNet-1}} & \textbf{52.86}                                & \textbf{47.46}        & \textbf{46.25}  & \textbf{13.08}                      & \textbf{26.42} & \textbf{46.49} & \textbf{23.55} & \textbf{32.08} & \textbf{26.00} & \textbf{25.90}  & \textbf{48.26}       
\\ \hline
\multicolumn{2}{|l|}{{\textbf{AffectiveNet-2}}} & {\textbf{61.20}}                                & {\textbf{58.12}}        & \textbf{-}  & \textbf{-}                      & \textbf{-} & \textbf{-} & \textbf{-} & \textbf{-} & \textbf{-} & \textbf{-}  & \textbf{-}       
\\ \hline
\end{tabular}
\end{table*}

\section{Experimental Results and Analysis}
\subsection{Database Prepossessing}
To test the effectiveness of AffectiveNet, we have utilized four benchmark datasets: CASME-I \cite{casme-2}, CASME-II \cite{casme-2}, CASME$^2$ \cite{camse-II} and SAMM \cite{SAMM}. These datasets are prepared to analyze the candid expressions under various challenges like illumination variations, subjects with different artifacts, ethnicity variations, age differences, gender inequalities etc.
\subsubsection{CASME-I}
The Chinese Academy of Sciences Micro-expression (CASME) \cite{casme-2} dataset comprises of 19 participants' spontaneous micro-expressions. The dataset samples are labeled with eight emotion classes as: contempt, tense, disgust, happiness, surprise, fear, sadness and repression with onset, peak and offset frame tags. However, in CASME-I dataset, some emotions like fear, sadness and contempt include very few samples and some of the emotion labels are ambiguous. Thus, most of the existing approaches \cite{stLbp, fdm} dropped these emotion classes to balance the inequality issue in datasets.  Recently, some methods \cite{tics,fhofo} have created new emotion classes by merging the existing emotions as positive, negative, surprise and other. In our experimental setup we have utilized the merged emotion classes and finally gathered 187 affective motion images as: positive: 9, negative: 50, surprise: 21 and others: 106. 
\begin{figure*}[!t]
\centering
	\includegraphics[width=\linewidth, height=3.0in]{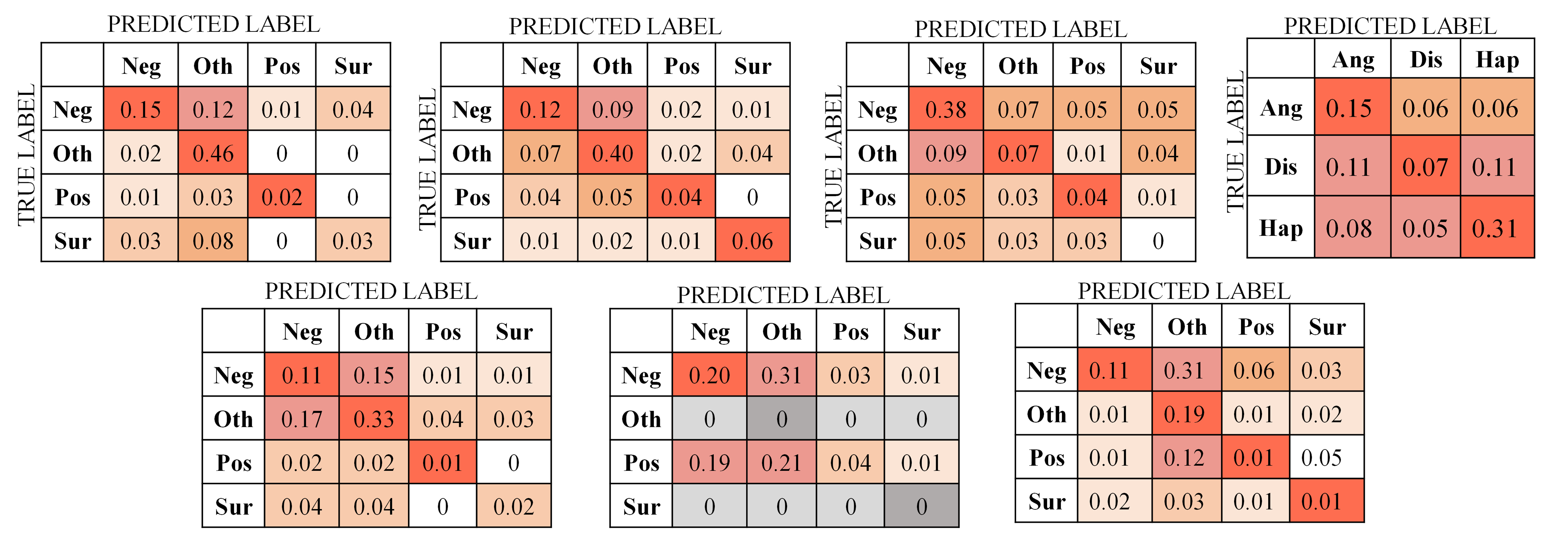}
	\vspace{-3mm}
	\caption{Confusion matrices of AffectiveNet for 4-class expression classification for a) CASME-I, b) CASME-II, c) SAMM d) CASME$^{2}$ and for e) CII2CI, f) CII2C$^{2}$, g) CII2S, in PIE and CDE setups, respectively.}
	\label{fig:Figure5}
\end{figure*}
\subsubsection{CASME-II}
The CASME-II \cite{casme-2} elicits 26 participants' micro expressions in a well-arranged laboratory with normal lighting to avoid the problem of illumination variation. Each frame is annotated with one of seven emotions as disgust, fear, happiness, other, repression, sadness and surprise. Similar to CASME-I, we have also converted CASME-II dataset into four categories and collected 251 affective-motion images as positive: 31, negative: 72, surprise: 25 and others: 126.
\subsubsection{CASME$^{2}$}
The CASME$^2$ \cite{camse-II} includes 22 subjects’ (6 female and 16 male) expressions captured at 30 fps with $640\times480$ resolution. The dataset has been annotated with three emotion classes: happy, anger and disgust based on AUs, self-decision of participants and emotion-evoking videos. In our experimental setup, we have selected a total 345 image sequences: anger-102, happy-155 and disgust-88 of micro expressions.
\subsubsection{SAMM}
The SAMM \cite{SAMM} dataset include 159 micro-expressions of 29 subjects with the largest ethnicity variations. The dataset is labeled with eight identified categories of expressions$:$ other, disgust, happiness, contempt, fear, sadness, surprise and anger. Similar to CASME-I and CASME-II, SAMM dataset also holds unequal data samples in emotion classes, thus we can combine emotion classes and compile 159 affective-motion images as positive: 26, negative:75, surprise:15 and others:43.
\subsection{Experimental Setup}
To evaluate the performance of the proposed method, we have chosen two sets of experiments: Person independent experiments (PIE) and Cross dataset experiments (CDE).
\subsubsection{Person independent experiments}
In literature \cite{fdm,spotrcnn,ststsnet}, mainly two types of evaluation techniques are used to validate the efficiency of MER systems: leave one video out (LOVO) and leave one subject out (LOSO) cross validation. In LOVO one expression video is used as testing set and remaining all videos are used for training set. Therefore, LOVO is evaluating the performance of MER in person dependent manner. Thus, LOVO is prone to subject biasing and does not validate the performance of system in effective manner. Thus, in this paper, all experiments are computed by using LOSO strategy. In LOSO, only one subject’s expressions are involved in testing set and remaining all subject’s expressions are used for training. This ensures robustness to unseen faces for expression recognition.
\subsubsection{Cross dataset experiments}
In this paper we have utilized the cross-dataset experiments (CDE) setup to evaluate the robustness and learnability of the AffectiveNet in cross domain. In CDE setup a dataset is used to train a model and other dataset are used as test set. In CDE, a different set of experiments are performed as follows. CI2CII: where CASME-I dataset is used as training set and CASME-II is testing set. Similarly, CI2C2, CI2S, CII2CI, CII2C2, CII2S, S2CI, S2CII and S2C2 experiments are conducted for other dataset combinations. Moreover, the Performance of proposed method is measured using recognition accuracy calculated by using Eq. \ref{eq:21}.
\begin{equation}\label{eq:21}
    Recog. Acc.=\frac{Total \, no.\, of\, correctly\, predicted\, samples}{Total\, no.\, of \,samples} \times 100
\end{equation}
\begin{figure*}[!t]
\centering
	\includegraphics[width=\linewidth, height=4.0in]{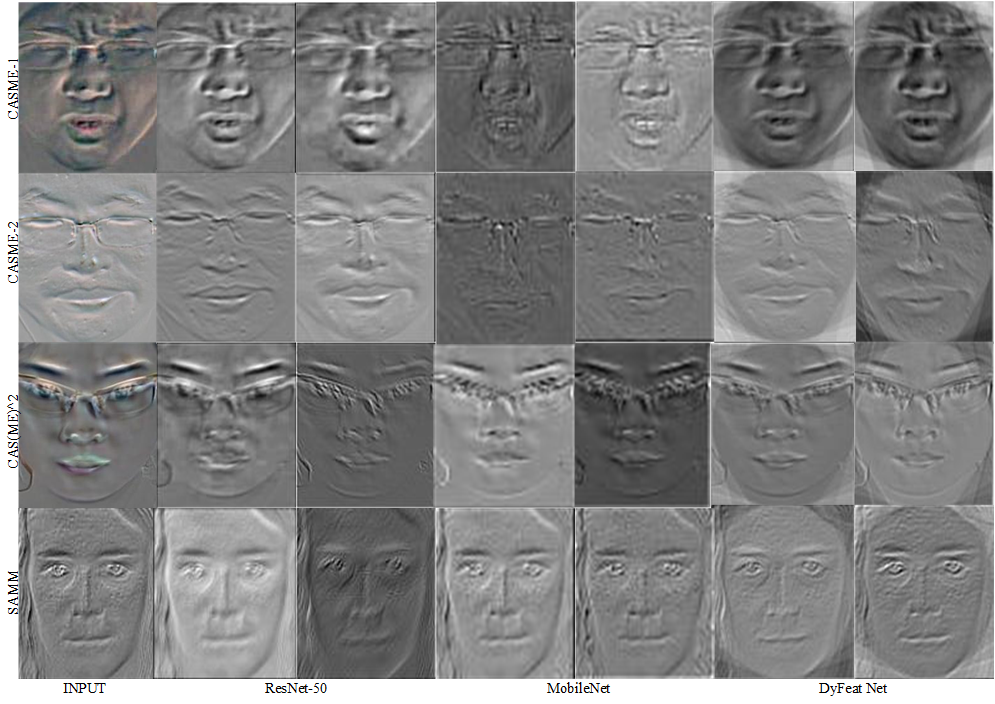}
	\vspace{-3mm}
	\caption{Qualitative representation of feature maps generated by existing networks: ResNet, MobileNet and proposed network over different micro expressions of four datasets: a) CASME-I: b) CASME-II: c) CASME$^2$ and d) SAMM. The red blocks validate that AffectiveNet is able to capture the furrow lines more accurately as compared to ResNet and MobileNet.}
	\label{fig:Figure6}
\end{figure*}

Furthermore, to examine the effectiveness of AffectiveNet, we have compared the proposed method with existing MER approaches by following two schemes.
\begin{enumerate}
\item We trained existing conventional networks: ResNet \cite{resnet}, MobleNet \cite{mobilenet}  by utilizing pre-trained weights over our experimental setup that ensure a fair comparison between state-of-the-art and proposed method. However, in case of other MER approaches \cite{tics,fhofo} we quoted the published results directly we follow the similar experimental setup.
\item Since, some of the recent approaches  \cite{microlstm,spontenous,spotrcnn} and \cite{ststsnet} etc. are follow contrast experimental setup in terms of total number of samples, participants, expression classes etc. or  dropped some of emotion classes due to a smaller number of images. Therefore to validate the effectiveness of proposed AffectiveNet, we have compared with the existing state-of-the-art approaches by following the experimental setup added in the  \cite{spotrcnn}.
\end{enumerate}

However, In our experiments, we have augmented the generated affective-motion images and create a large pool of data to avoid the problem of over-fitting in training. Moreover, to train the network we have used SGD optimizer and SoftMax loss function with 10$^-$3 learning rate.
\subsection{Quantitative Analysis}
This section provides a comparative analysis of the obtained accuracy rates between the existing and proposed network for both PIE and CDE experiments.
\subsubsection{Person independent experiments}
Recognition accuracy results over CASME-I, CASME-II, CASME$^2$ and SAMM datasets for existing state-of-art and affective approaches for PIE setup are tabulated in Table-\ref{tab:Table1}, respectively. Specifically, for CASME-I, Proposed Network secures 33.22\%, and 41.95\% more accuracy as compared to MobileNet and ResNet respectively. Moreover, AffectiveNet also outperforms the existing handcrafted MER: TICS and FHOFO approaches by 5.13\% and 1.00\%, respectively. For CASME-II, our network achieves 31.33\% and 29.46\% more accuracy as compared to MobileNet and ResNet respectively. Furthermore, proposed model yields 0.47\% and 5.72\% better accuracy rates as compared to TICS and FHOFO respectively.  For CASME$^2$ dataset, our proposed method secures 17.81\% and 6.28\% improvement over MobileNet and ResNet respectively. Similarly, for SAMM dataset, AffectiveNet outperforms the existing approaches MobileNet and ResNet by 7.26\% and 10.96\% accuracy rates, respectively. 
\subsubsection{Cross dataset experiments}
Comparative analysis results of the conventional CNN network and proposed network for CDE setup are tabulated in Table-\ref{tab:Table2}. From the Table-\ref{tab:Table2}, it is clear that, proposed model  outperforms CDE experiment results and validates the strength of proposed network. Particularly, AffectiveNet gains 35.58\%, 0.58\%, 15.73\% and 34.39\%, 9.08\%, 11.96\% more accuracy for CI2CII, CI2C2, CI2S setups as compared to ResNet and MobileNet, respectively. Moreover, AffectiveNet yields 35.68\%, 0.01\%, 23.27\% and 22.71\%, 17.15\%, 13.84\% better accuracy rates for CII2CI, CII2C$^2$ and CII2S experiments compared to ResNet and MobileNet respectively. Similarly, for S2CI, S2CII and S2C$^2$  proposed model attains 5.85\%, 9.3\%, 19.2\% and 22.71\%, 13.26\%, 40.25\% as compared to ResNet and MobileNet, respectively.

To analyze class-wise recognition accuracy (true positives and false positives), we have presented the confusion matrices for all PIE and CDE experiments in Fig. \ref{fig:Figure5}.
\subsection{Qualitative Analysis}
The learning capability of proposed network is compared with the state-of-the-art networks are shown in Fig. \ref{fig:Figure6}. Fig. \ref{fig:Figure6}. demonstrates the three most effective visual representations of different emotion classes as CASME-I: disgust, CASME-II: Happy, CASME$^2$: disgust and SAMM: Others. From figure, it is clear that the response feature maps significantly assist in preserving the dynamic variations in different expressive regions of the facial image. For example, in Disgust: eyes, eyebrows, mouth regions; in happy: eyebrows, lips, mouth and in others: eyes, mouth; give maximum affective response for related facial expressions. Therefore, we conclude that AffectiveNet preserves more relevant feature responses to outperform the existing CNN based networks ResNet-50 and MobileNet for almost all emotion classes. 
\subsection{Computational Complexity}
This section provides a comparative analysis of the computational complexity between the existing and proposed networks. The total number of parameters involved in each network are tabulated in Table-\ref{tab:Table3}. The proposed AffectiveNet has only 2.3 million learnable parameters which are very less as compared to other existing benchmark models like: MobileNet: 3.2M, VGG-16: 138M, VGG-19: 144M and ResNet: 11.7M. Moreover, proposed network architecture has fewer depth channels and hidden layers as compared to former methods. Furthermore, AffectiveNet takes only 8.3 MB memory storage which is very less as compared to MobileNet: 25.3 MB, VGG-16: 515, VGG-19: 535 and ResNet: 44 MB.

\begin{figure}[!t]
\centering
	\includegraphics[width=\linewidth, height=3.5in]{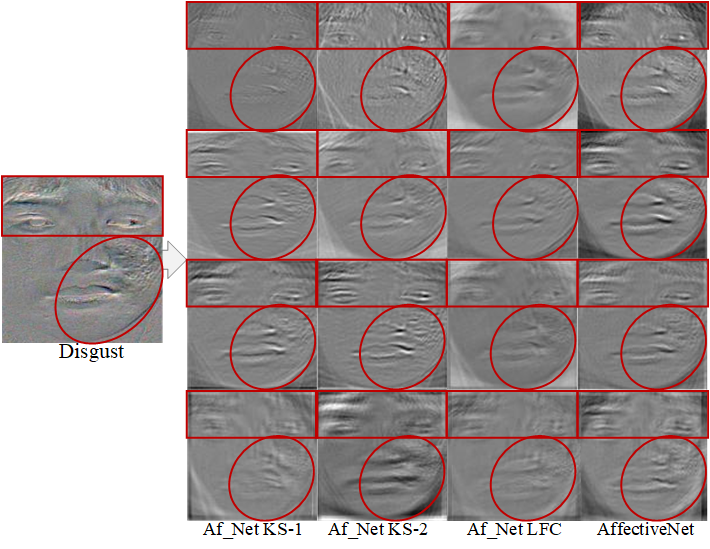}
	\vspace{-3mm}
	\caption{The neuron visualization of responses for disgust emotion captured at 1st multiscale CNN layers of ablation experiments a) Af-Net-KS-1 b) Af-Net-KS-2, c) Af-Net-LFC and proposed AffectiveNet.}
	\label{fig:Figure7}
\end{figure}
\begin{table}[]
\centering
\caption{Computational Complexity Analysis of AffectiveNet and Exitsing Networks.}
\label{tab:Table3}
\vspace{-3mm}
\begin{tabular}{cccc}\hline
Network    & \begin{tabular}[c]{@{}l@{}}  $\#$ Parameters \\(in millions)\end{tabular} & \begin{tabular}[c]{@{}l@{}} $\#$ Memory \\ (in megabytes)\end{tabular} &  \begin{tabular}[c]{@{}l@{}} Speed \\ (in seconds)\end{tabular}\\
\hline \hline
VGG-16\cite{vgg} &   138  &     515    & {17.8} \\
VGG-19 \cite{vgg} & 144 & 535 & {21.2} \\
ResNet -50 \cite{resnet} & 26 & 44 & 19\\
MobileNet \cite{mobilenet} & 4.2 & 25.3 & 12 \\
\textbf{Af-Net-KS-1} & \textbf{2.3} & \textbf{8.5} & \textbf{8.5} \\
\textbf{Af-Net-KS-2} & \textbf{2.5} & \textbf{9.4} & \textbf{8.6} \\
\textbf{Af-Net-LFC} & \textbf{1.1} & \textbf{4.0} & \textbf{8.6} \\
{\textbf{Af-Net-WoMFL}} & {\textbf{1.0}} & {\textbf{3.4}} & {\textbf{4.5}} \\
{\textbf{Af-Net-$3\times3$}} & {\textbf{2.1}} & {\textbf{7.8}} & {\textbf{5.4}} \\
{\textbf{Af-Net-$1\times1$}} & {\textbf{2.1}} & {\textbf{8.1}} & {\textbf{5.1}} \\
\textbf{AfffectiveNet-1} & \textbf{2.2} & \textbf{8.3} & \textbf{8.7}\\\hline
\end{tabular}
\end{table}

\subsection{Ablation Study}
In order to investigate the deep insights of AffectiveNet, we have conducted six more supplementary experiments for ablation study as represented in Table-\ref{tab:Table1}. This section mainly focuses on examining the effect of different kernel sizes in EncapFeat block,linearly connected fully connected layer in MFL block, effect of different filter sizes and MFL block. First, we have examined the impact of two large kernel sizes like $\left(5\times5,7\times7\right)$   and $\left(7\times7,11\times11\right)$instead of  $\left(3\times3,5\times5\right)$ in EncapFeat block named as Af-Net-KS-1. and Af-Net-KS-2, respectively. Therefore, we have observed that smaller size kernels are more preferable for micro-expression recognition. Kernels with large scale ignore the minute transitional information which is quite important in micro expression. From the Fig. 3 and 7 it is clear that kernel size $\left(11\times11\right)$ skips the micro edge variations and preserved only high-level edges.  Second, we have analyzed the effect of linearly connected FC layers in MSF blocks in proposed method, named as Af-Net-LFC. Af-Net-LFC fails to learns the pertinent features and degrades the performance of network. Third, we computed results by dropping the MFL block named as af-Net-WoMFL to investigate the role of MFL block in learning of dicriminative variations of micro expressions.  Further, we have examined the effect of multi-scale filters by replacing all filters with $\left(3\times3\right)$ named as af-Net-$\left(3\times3\right)$. Finally, to analyse the effects of $\left(1\times1\right)$ sized filters, we execute the AffectiveNet by replacing $\left(3\times3\right)$ with $\left(1\times1\right)$ in MICRoFeat block.Quantitative results, represents in Table-\ref{tab:Table1} validates the performance of AffectiveNet over other supplementary results. Moreover. Thus, by observing ablation studies experimental results, we can conclude that our proposed model has generated best results as compared to other combinations.
\section{Conclusion}
This paper presents an AffectiveNet: affective-motion feature learning for micro-expression recognition. First, we computed single instance responses of the affective-motion images from micro expression sequences which preserves the facial movements into one instance. Further, the generated single instance is processed through the AffectiveNet to estimate the networks performance. In AffectiveNet two blocks are introduced MICRoFeat and MFL, to learn the micro expression features. MICRoFeat block holds multi-scale filters as $3\times3$, $5 \times5$,$ 7\times 7$ and $11 \times 11$ to extract the comprehensive and detailed edge variations from affective images. Thus, MICRoFeat block is responsible to capture edge variations from small regions to extensive regions. While, MFL block incorporats two-stage FC layers to more discriminative features of micro expressions and allows network to define the disparities between emotion classes. Moreover, the proposed network has a small number of parameters that reduce the training and testing time of the MER system. The effectiveness of system is evaluated on the benchmark dataset CASME-I, CASME-II, CASME$^2$ and SAMM. It is evident from experimental results, visual demonstration, complexity analysis and ablation study that AffectiveNet has achieved better accuracy rates as compared to state-of-the-art approaches for MER.

\section*{Acknowledgment}

This work was supported by the Science and Engineering Research Board (under the Department of Science and Technology, Govt. of India) project $\#$SERB/F/9507/2017. The authors would like to thank our Vision Intelligence lab group for their valuable support. We are also thankful to Shastri Indo-Canadian Institute for their support in the form of SRS fellowship.

\vskip -2pt plus -1fil
\begin{IEEEbiography}[{\includegraphics[width=1in,height=1.50in,clip,keepaspectratio]{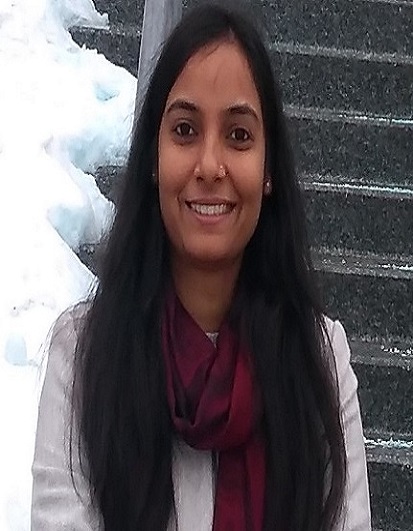}}]%
{Monu Verma}
received her B. Tech degree in Computer Science and Engineering from GEC Bikaner, India, in 2013. She received her M. tech degree in 2016 from NIT Jalandhar, India. She is currently pursuing her Ph.D. with the Department of Computer Science and Engineering, MNIT Jaipur, India. She is a life member of Vision Intelligence Lab  @ MNIT, Jaipur. Her research interests include facial expression recognition, depression analysis, micro expression recognition,  hand posture classification and finger sign analysis, .
\end{IEEEbiography}
\vskip -2pt plus -1fil
\begin{IEEEbiography}[{\includegraphics[width=1in,height=1.50in,clip,keepaspectratio]{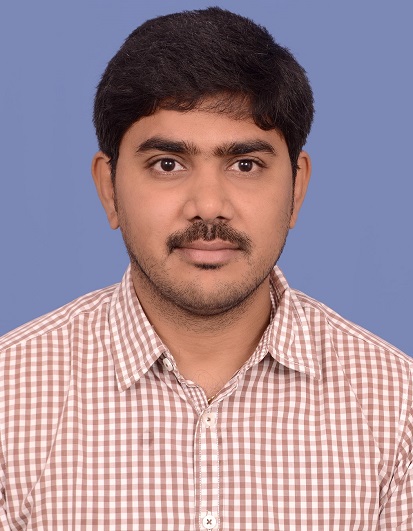}}]%
{Santosh Kumar Vipparthi}
 received his B.E. degree in Electrical and Electronics Engineering from Andhra University, India. Further, he received his M. Tech. and Ph. D. in Systems Engineering from IIT BHU, Varanasi, India. Currently, he is an assistant professor in the Department of Computer Science and Engineering, MNIT Jaipur, India. He leads Vision Intelligence Lab  @ MNIT with research focused important visual perception tasks such as object detection, human emotion recognition, aberrant event detection, image retrieval, Gesture recognition, Motion analysis, etc.
\end{IEEEbiography}
\vskip -2pt plus -1fil
\begin{IEEEbiography}[{\includegraphics[width=1in,height=1.50in,clip,keepaspectratio]{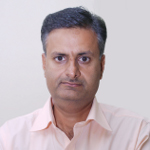}}]%
{Girdhari Singh}
 received the B.E. degree in Computer Engineering from Amravati University, Maharastra India, in 1990. Afterwards, he received his MS in Software Engineering from BITS Pilani, India in 1996. Further, He received his Ph.D. in Computer Engineering from MNIT, Jaipur, India in 2009. Currently, he is working as an associate professor in the department of computer science and engineering, MNIT, Jaipur, Rajasthan, India. His major fields of research are software engineering, intelligent systems image processing and machine learning.
\end{IEEEbiography}

\end{document}